\documentclass[aps,prl,twocolumn,floatfix]{revtex4}
\usepackage{amssymb}
\usepackage{amsmath}
\usepackage{graphicx}

\begin{document}

\title{Analysis of Dirac and Weyl points in topological semimetals via oscillation effects}

\author{G.~P.~Mikitik}
\affiliation{B.~Verkin Institute for Low Temperature Physics \&
Engineering, Ukrainian Academy of Sciences, Kharkov 61103,
Ukraine}

\author{Yu.~V.~Sharlai}
\affiliation{B.~Verkin Institute for Low Temperature Physics \&
Engineering, Ukrainian Academy of Sciences, Kharkov 61103,
Ukraine}

\begin{abstract}
We calculate the extremal cross sectional areas and cyclotron masses for the Fermi-surface pockets in Dirac and Weyl topological semimetals. The calculation is carried out for the most general form of the electron energy bands in the vicinity of the Weyl and Dirac points. Using the obtained formulas, one can find parameters characterizing the Dirac and Weyl electrons in the topological semimetals from appropriate experimental data. As an example, we consider the W1 electrons in TaAs.
\end{abstract}

\maketitle

The topological Weyl and Dirac semimetals have attracted much attention in recent years; see, e.g., reviews \cite{armit,bernevig,gao,wang-r,m-sh19} and references therein. In the Weyl semimetals, two electron bands contact at discrete (Weyl) points of the Brillouin zone and disperse linearly in all directions around these specific points. The same type of the band contact occurs in the Dirac semimetals, but the bands are double degenerate in spin. In other words, a Dirac point can be considered as a superposition of two Weyl points in the quasi-momentum space. The chemical potential of electrons, $\zeta$, in the Weyl and Dirac semimetals is close to the band-contact energy $\varepsilon_d$.
A number of the Dirac and Weyl semimetals  were discovered in recent years \cite{armit,bernevig,zhang-nc19}.

Various oscillation effects are widely used in experimental investigations of the topological semimetals. In particular,
measurements of the quantum-oscillations phase related to the so-called Berry phase \cite{prl} were carried out in a host of works in order to detect the Weyl and Dirac electrons in semimetals, see, e.g., review \cite{m-sh19} and references therein. Beside this,
using the Shubnikov - de Haas and de Haas - van Alphen effects, the extremal cross sectional areas of the Fermi surfaces and the cyclotron masses corresponding to these cross sections were measured for a number of the Weyl \cite{huang15,luo15,shek,hu16,du16,sergelius,wang16,arnold16} and Dirac \cite{he-r(h),pari,liang,zhao15,nara15,xiang15,desr,cao15,he16,crassee}
semimetals. In this paper we present formulas for such areas and masses. These formulas will allow one to obtain the parameters characterizing the Dirac and Weyl points in the topological semimetals from the experimental data.

\begin{figure}[tbp] 
 \centering  \vspace{+9 pt}
\includegraphics[scale=.90]{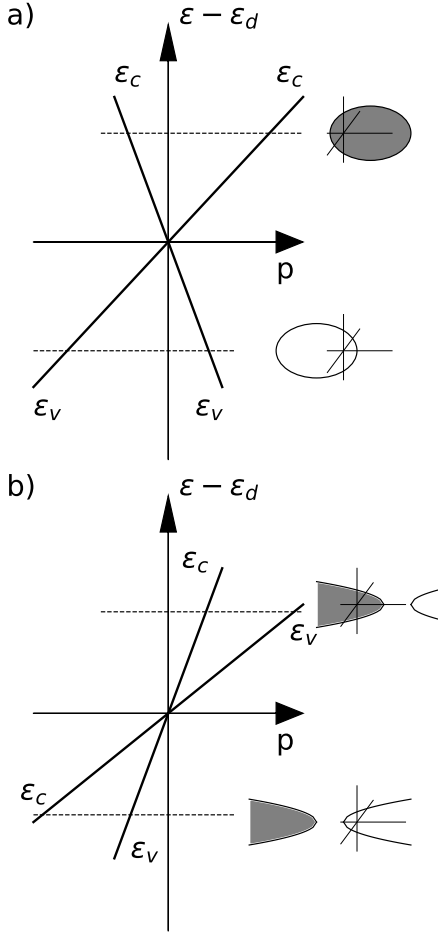}
\caption{\label{fig1} Dispersion relations $\varepsilon_c(p)$ and $\varepsilon_v(p)$ of the two contacting bands in the vicinity of a Weyl (Dirac) point in the cases of $\tilde a^2<1$ (a) and $\tilde a^2>1$ (b). On the right, the Fermi surfaces at $\zeta-\varepsilon_d<0$ and $\zeta-\varepsilon_d>0$ are shown together with the Weyl (Dirac) point which is the origin of the coordinate axes. The shaded and white surfaces correspond to the electron and hole charge carriers, respectively.
 } \end{figure}   

The most general electron dispersion relations  $\varepsilon_{c,v}({\bf p})$ of the two bands $c,v$ in the vicinity of a Weyl (Dirac) point look as follows \cite{m-sv,m-sh,m-sh19}:
 \begin{eqnarray}\label{1}
 \varepsilon_{c,v}({\bf p})=\varepsilon_d+{\bf a}\cdot{\bf p}\pm E({\bf p}),
 \end{eqnarray}
where the quasi-momentum ${\bf p}$ is measured from this point, and $[E({\bf p})]^2$ is a positively definite quadratic form in the components of the vector ${\bf p}$. Below we shall choose the coordinate axes along principal directions of this form. In this case, one has
 \begin{eqnarray}\label{2}
[E({\bf p})]^2=b_{11}p_1^2+b_{22}p_2^2+b_{33}p_3^2,
 \end{eqnarray}
where  $b_{11}$, $b_{22}$, $b_{33}$ are the positive constants.
The scaling of the coordinate axes, $\tilde p_i=p_i\sqrt{b_{ii}}$, transforms Eqs.~(\ref{1}), (\ref{2}) into the form,
 \begin{eqnarray*}
 \varepsilon_{c,v}=\varepsilon_d+\tilde{\bf a}\cdot\tilde{\bf p}\pm |\tilde {\bf p}|,
 \end{eqnarray*}
that depends only on the constant dimensionless vector
 \[
 \tilde {\bf a}\equiv \left( \frac{a_1}{\sqrt{b_{11}}},\frac{a_2}{\sqrt{b_{22}}}, \frac{a_3}{\sqrt{b_{33}}}\right).
 \]
The vector $\tilde{\bf a}$ characterizes a tilt of the bands $\varepsilon_{c,v}({\bf p})$, and its length is the most important parameter of dispersion relation (\ref{1}). When  the length of $\tilde {\bf a}$ is less than unity,
\begin{eqnarray*}
\tilde a^2=\frac{a_1^2}{b_{11}}+ \frac{a_2^2}{b_{22}} +\frac{a_3^2}{b_{33}} < 1,
 \end{eqnarray*}
the dispersion relations $\varepsilon_{c,v}({\bf p})$ looks like in Fig.~1a. In this case, the Fermi surface is either a closed hole pocket if  $\zeta<\varepsilon_d$ or a closed electron pocket if $\zeta > \varepsilon_d$. When $\tilde a^2>1$, there is a direction in the ${\bf p}$-space along which the dispersion relations $\varepsilon_{c,v}({\bf p})$ look like in Fig.~1b, and ``open'' electron and hole pockets of the Fermi surface exist both at $\zeta < \varepsilon_d$ and $\zeta > \varepsilon_d$. It is necessary to emphasize that the parameter $\tilde a^2$, which specifies the tilt of the bands,  differs from zero for all the Weyl points and for the Dirac points induced by the band inversion \cite{armit} since all these points do not belong to the class of highly-symmetric points in the Brillouin zone of the topological semimetals. If $\tilde a^2<1$, a Weyl (Dirac) semimetal falls into the type I, whereas the case $\tilde a^2>1$ corresponds to the so-called type-II Weyl (Dirac) semimetals \cite{sol}. Below we consider only the type-I semimetals which have the closed Fermi surfaces.

At small $|\zeta-\varepsilon_d|$, the Fermi surfaces near the Weyl (Dirac) points are ellipsoids, with the center of the ellipsoids being displaced from these points (i.e., from  ${\bf p}=0$) by the vector that is proportional to $(\zeta-\varepsilon_d)$. Beside this, if at least two components of $\tilde{\bf a}$ differ from zero, the axes of the ellipsoid deviate from the axes of the coordinate system. The displacement of the Fermi surface  leads to the fact that its maximal cross section perpendicular to a unit vector ${\bf n}$  generally does not pass through the Weyl (Dirac) point ${\bf p}=0$, Fig.~\ref{fig1}. Using the dispersion relation (\ref{1}), (\ref{2}), one can calculate both the maximal cross sectional area $S_{\rm max}$ of the Fermi surface at an arbitrary direction ${\bf n}$ of the magnetic field $H$ and the cyclotron mass $m_*=(1/2\pi)(\partial S_{\rm max}/\partial \zeta)$ corresponding to this cross section,
\begin{eqnarray}\label{3}
S_{\rm max}&=&\frac{\pi(\zeta-\varepsilon_d)^2}{R_n^{1/2}(1-\tilde a^2)}, \\
m_*&=&\frac{(\zeta-\varepsilon_d)}{R_n^{1/2}(1-\tilde a^2)}. \label{4} \end{eqnarray}
The angular-dependent factor $R_n$  in these expressions has the form,
\begin{eqnarray}\label{5}
R_n= b_{11}b_{22}b_{33}[(1-\tilde a^2)\tilde{\bf n}^2+(\tilde {\bf a}\cdot \tilde{\bf n})^2]=\!\!\sum_{i,j=1}^{3}\kappa^{ij}n_in_j,
\end{eqnarray}
where
\begin{eqnarray*}
 \tilde {\bf n}&\equiv& \left(\frac{n_1}{\sqrt{b_{11}}}, \frac{n_2}{\sqrt{b_{22}}}, \frac{n_3}{\sqrt{b_{33}}}\right), \\
 \kappa^{ij}\!\!&=&\frac{b_{11}b_{22}b_{33}}{(b_{ii}b_{jj})^{1/2}} \left[(1-\tilde a^2)\delta_{ij}+\tilde a_i\tilde a_j\right],
 \end{eqnarray*}
and $\delta_{ik}$ is the Kronecker symbol.

It follows from Eqs.~(\ref{3}) and (\ref{4}) that
 \begin{eqnarray}\label{6}
|\zeta - \varepsilon_d|=\frac{S_{\rm max}}{ \pi |m_*|}=\frac{2e\hbar F}{c|m_*|},
 \end{eqnarray}
where $F$ is the frequency of the quantum oscillations produced by the cross-sectional area  $S_{\rm max}$ in a physical quantity $Q$ [i.e., the first harmonic of $Q$ is proportional to $\cos(2\pi F/H+ \phi_0)$ where $\phi_0$ is some phase]. Therefore, if the frequency $F$ and the cyclotron mass $m_*$ have been measured at least for one direction of the magnetic field, formula (\ref{6}) enables one to find the position of the chemical potential $\zeta$ relative to the energy $\varepsilon_d$ of the Weyl (Dirac) point.

The dispersion relation (\ref{1}), (\ref{2}) is determined by the six parameters: $b_{11}$, $b_{22}$, $b_{33}$, $\tilde a_1$, $\tilde a_2$, $\tilde a_3$. Beside this, the orientation of the principal axes of the quadratic form $[E({\bf p})]^2$ relative to the crystallographic axes of the semimetal can be described by three angles, and hence the nine parameters define a Weyl (Dirac) point in the general case. The angular dependences of the frequency $F$ are specified by the factor $1/\sqrt{R_n}$ in Eq.~(\ref{3}), and this factor is defined by the six constants $\kappa^{ij}$. Hence, an approximation of experimental angular dependences of this frequency with formulas (\ref{3}), (\ref{5}) together with Eq.~(\ref{6}) provides possibility to determine the six combinations of the parameters  characterizing the dispersion relation.

The densities $n_W$ and $n_D$ of the Weyl and Dirac charge carriers can be expressed in terms of directly-measurable frequencies of the quantum oscillations,
\begin{eqnarray}\label{7}
n_W=\frac{N_W V}{(2\pi\hbar)^3},\ \ \ \ \ \ n_D=\frac{2N_D V}{(2\pi \hbar)^3},
 \end{eqnarray}
where $V$ is the volume of a Weyl or Dirac pocket in the Brillouin zone,
 \begin{eqnarray}\label{8}
 V\!\!=\!\!\frac{4 [S_{\rm max}^{(1)}S_{\rm max}^{(2)}S_{\rm max}^{(3)}]^{1/2}}{3\pi^{1/2}}\!=\!\frac{8\sqrt{2}\pi (e\hbar)^{3/2}(F_1F_2F_3)^{1/2}}{3c^{3/2}},
 \end{eqnarray}
$N_W$ and $N_D$ are the numbers of the equivalent pockets, $F_1$ and $F_3$ are the maximal and minimal frequencies produced by the pocket when the magnetic field rotates in various planes, and $F_2$ corresponds to the direction of ${\bf H}$ perpendicular to the directions at which $F_1$ and $F_2$ occur. The cross-sectional areas $S_{\rm max}^{(i)}$ correspond to the frequencies $F_i$, and these cross sections are mutually orthogonal.

Consider now two special cases in more detail.

\subsection{Dirac point}

In the  Dirac semimetals induced by the band inversion, the Dirac points can lie only in symmetry axes of the third, fourth or sixth order \cite{armit}. The well-known Dirac semimetals Na$_3$Bi and Cd$_3$As$_2$ just fall into this class. In this case, one of the principal axes of $[E({\bf p})]^2$ coincides with the symmetry axis, which we designate as the axis $3$. The symmetry also imposes the  restrictions: $b_{11}=b_{22}=b_{\perp}$ (generally $b_{\perp}\neq b_{33}$), ${\bf a}=(0,0,a_3)$. With these restrictions, formulas (\ref{3}) and (\ref{5}) give the following expressions for $S_{\rm max}(\theta)$, the maximal area of the cross section that is perpendicular to the magnetic field tilted at the angle $\theta$ to the symmetry axis,
\begin{eqnarray}\label{9}
S_{\rm max}(0)&=&\frac{\pi(\zeta-\varepsilon_d)^2}{b_{\perp}(1-\tilde a_3^2)}, \\
\frac{S_{\rm max}(\theta)}{S_{\rm max}(0)}&=&\frac{1}{\sqrt{\cos^2\theta+\epsilon^2 \sin^2\theta}},
 \label{10}
 \end{eqnarray}
where $\epsilon^2=(1-\tilde a_3^2)b_{33}/b_{\perp}$. Thus, if  $S_{\rm max}(0)$, $m_*(0)$, $S_{\rm max}(\pi/2)$ are measured, one can find $|\zeta-\varepsilon_d|$, $b_{\perp}(1-\tilde a_3^2)$, and $\epsilon$ with formulas (\ref{6}), (\ref{9}), (\ref{10}). Note that dependence (\ref{10}) has the standard form typical of an ellipsoidal Fermi surface. However, the anisotropy of the Fermi surface $\epsilon= S_{\rm max}(0)/S_{\rm max}(\pi/2)$ contains the factor $\sqrt{1-\tilde a_3^2}$ which is caused by the tilt of the bands $\varepsilon_{c,v}({\bf p})$. The density $n_D$ of the Dirac charge carriers can be found with Eqs.~(\ref{7}), (\ref{8}) where $S_{\rm max}^{(1)}S_{\rm max}^{(2)}S_{\rm max}^{(3)}=S_{\rm max}(0)[S_{\rm max}(\pi/2)]^2$ now.

\subsection{Weyl point near a reflection plane}

Consider a Weyl point for which the parameters meet the following restrictions: $b_{11}, b_{22} \gg b_{33}$ and $a_3\ll a_1$, but $\tilde a_3\equiv a_3/\sqrt{b_{33}}$ can have any value satisfying the condition $(\tilde a_3)^2 <1-(\tilde a_1)^2-(\tilde a_2)^2$. Such a point may appear if it results from a nodal line that lies in the reflection plane $1-3$ of the crystal without the inversion symmetry. This line exists in neglect of the spin-orbit interaction. In this case, for any point of the line, the vector ${\bf a}$ lies in the reflection plane (i.e., $a_2=0$), whereas one of the local values of $b_{11}$, $b_{33}$ is equal to zero (for definiteness, let $b_{33}=0$) \cite{m-sv,m-sh,m-sh19}.  A nonzero strength of spin-orbit  interaction lifts the degeneracy of the electron bands along the nodal line and can lead to the appearance of two Weyl points disposed near the reflection plane (symmetrically relative to it) \cite{arnold16}. If the spin-orbit interaction does give rise to the Weyl point {\em slightly} displaced from the plane, one may expect that $a_1$, $a_3$, $b_{11}$, $b_{33}$ will experience small changes, and the condition $b_{11}, b_{22} \gg b_{33}$ will hold true for the point. The fact of the appearance of the closed Fermi pocket surrounding the Weyl point provides the fulfilment of the condition $(\tilde a_3)^2 <1-(\tilde a_1)^2-(\tilde a_2)^2$ which means that such a pocket can occur near the point of the line where $a_3$ is relatively small, $a_3\lesssim \sqrt{b_{33}}\ll \sqrt{b_{11}}\sim a_1$. Although one may also expect that $\tilde a_2\approx 0$ for the Weyl point, the closeness of the two Weyl points to each other may noticeably modify the values of $b_{22}$ and $\tilde a_2$, and so we do not impose any restriction on these parameters.

\begin{figure}[tbp] 
 \centering  \vspace{+9 pt}
\includegraphics[scale=.50]{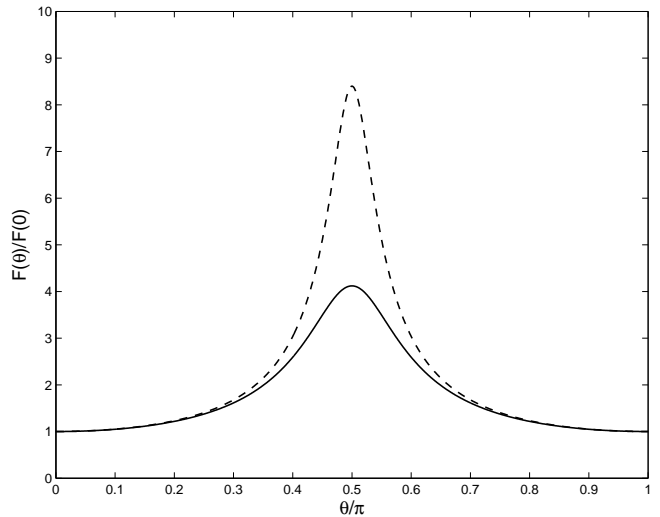}
\caption{\label{fig2} The frequency of quantum oscillations, $F=cS_{\rm max}/(2\pi e\hbar)$, versus the angle $\theta$ between the direction of the magnetic field and the axis $3$, Eqs.~(\ref{3}) and (\ref{5}). Here $\tilde a_1=0.5$ $\tilde a_2=0.47$, $\tilde a_3=0$, $b_{22}/b_{11}=4$, $b_{33}/b_{11}=0.04$ (these values correspond to the first set of the parameters in Table I). The angle $\theta$ changes either in the $1-3$ plane (the solid line) or in the $2-3$ plane (the dashed line).
 } \end{figure}   

For the Weyl point that results from the nodal line, the perpendicular to this line in the reflection plane, the normal to the plane, and the direction along the line are close to the directions of the axes $1, 2$ and $3$, respectively. In Fig.~\ref{fig2} we show the angular dependences of the frequency  $F(\theta)=cS_{\rm max}(\theta)/(2\pi \hbar e)$ where $\theta$ is the angle between the magnetic field and the axis $3$. This angle changes either in the $1-3$ plane or in the $2-3$ plane. Since in a tetragonal crystal like TaAs, the equivalent Weyl pockets exist near the two perpendicular reflection planes, both the dependences  presented in Fig.~\ref{fig2} can be  observed  for $H$ rotating in each of the planes. In particular, the dependences shown in Fig.~\ref{fig2} are similar to those found for the so-called W1 electrons in TaAs; see Fig.~3a in Ref.~\cite{arnold16}. In Fig.~\ref{fig2} we take $\tilde a_3=0$. If $\tilde a_3\neq 0$, the minimum value of $F(\theta)$ is reached at the nonzero angle $\theta_m$ in the plane $1-3$,
 \begin{equation}\label{11}
\theta_m\approx \frac{\tilde a_1 \tilde a_3}{1-\tilde a_1^2-\tilde a_2^2}\sqrt{\frac{b_{33}}{b_{11}}}.
 \end{equation}
A similar formula describes the position of the minimum of  $F(\theta)$ in the plane $2-3$ (in this case, $\tilde a_1$, $\tilde a_2$ and $b_{11}$ are replaced by $\tilde a_2$, $\tilde a_1$ and $b_{22}$, respectively). These nonzero $\theta_m$ are due to the above-mentioned deviation of the Fermi-surface axes from the coordinate axes.

\subsection{Example: W1 electrons in TaAs}

\begin{figure}[tbp] 
 \centering  \vspace{+9 pt}
\includegraphics[scale=1]{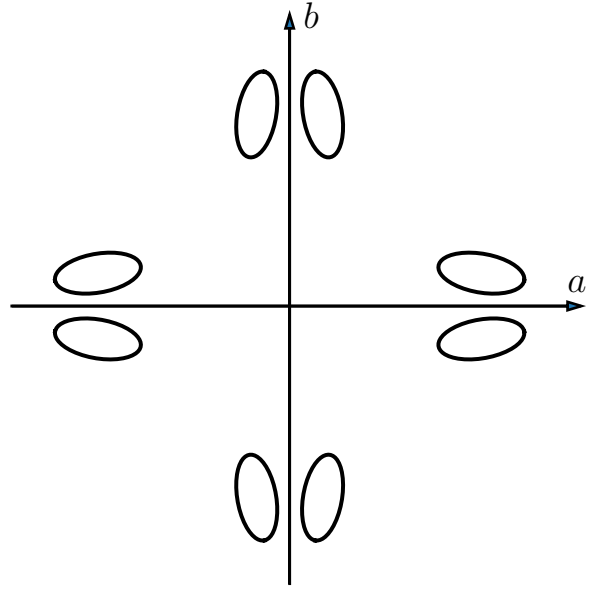}
\caption{\label{fig3} The outline of the cross sections of the W1 ellipsoids by the $a-b$ plane in TaAs.
 } \end{figure}

As an example, let us analyze the known experimental data for the W1 electrons in the Weyl semimetal TaAs \cite{arnold16}. Near the W1 points the appropriate nodal lines are parallel to the $c$ axis, and therefore it is reasonable to suppose that the directions of the axes $1, 2, 3$  coincide with the directions of the crystallographic axes $a, b, c$. Arnold et al. \cite{arnold16} found that the frequency of quantum oscillations $F$ changes like in Fig.~\ref{fig2} when the direction of the magnetic field varies in the reflection plane $c-a$ from the $c$ axis ($\theta=0$) to the $a$ axis  ($\theta=\pi/2$), and they obtained $F(0)\approx 7$ T, $m_*(0)/m \approx 0.057$. At $\theta\neq 0$, the frequency $F(\theta)$ splits into the two branches $F_a(\theta)$, $F_b(\theta)$ associated with the ellipsoids lying near the axes $a$ and $b$ (Fig.~\ref{fig3}), and the frequencies $F_a(\pi/2)$ and $F_b(\pi/2)$ take the values $29$ T and $59$ T. Using formula (\ref{6}) and the values of $F(0)$ and $m_*(0)$, we arrive at $\zeta-\varepsilon_d\approx 28.4$ meV. This result is close to the value $26$ meV obtained in the band-structure calculations \cite{arnold16}. Using Eq.~(\ref{6}), we can also predict the values of the cyclotron masses $m_*(\pi/2)/m \approx 0.24$  and $0.48$ which  correspond to the frequencies $29$ T and $59$ T, respectively.

With $\zeta-\varepsilon_d\approx 28.4$ meV and Eqs.~(\ref{3}), we find  $R_n^{1/2}(1-\tilde a^2)$ at $\theta=0$,
\begin{eqnarray*}
R_c^{1/2}(1-\tilde a^2)
= \frac{c(\zeta-\varepsilon_d)^2}{2e \hbar F(0)}\approx 8.75\cdot 10^{10}\,\frac{{\rm m}^2}{{\rm s}^2},
   \end{eqnarray*}
where $R_c\equiv R_n|_{\theta=0}$. On the other hand, Eq.~(\ref{5}) gives
\begin{eqnarray}\label{12}
R_c^{1/2}(1-\tilde a^2)\approx \sqrt{b_{11}b_{22}(1-\tilde a_1^2-\tilde a_2^2)}(1-\tilde a^2),
   \end{eqnarray}
and hence we have found the value of the right hand side of this expression.

According to Eq.~(\ref{5}), the factor $R_n^{1/2}$ at $\theta=\pi/2$  looks like
\begin{eqnarray*}
 \sqrt{R_a}&\approx &\sqrt{b_{22}b_{33}(1-\tilde a_2^2-\tilde a_3^2)}, \\
 \sqrt{R_b}&\approx &\sqrt{b_{11}b_{33}(1-\tilde a_1^2-\tilde a_3^2)},
   \end{eqnarray*}
for the W1 ellipsoids lying near the $a$ and $b$ axes, respectively. Since there are no visible displacements of the minima of $F_a(\theta)$ and $F_b(\theta)$ from the point $\theta=0$ in Fig~3a of Ref.~\cite{arnold16}, we conclude that the parameter $\tilde a_3$ is small for the W1 electrons, i.e., $\tilde a_3\approx 0$, and so $\tilde a^2\approx (\tilde a_1)^2+(\tilde a_2)^2$. Therefore, the ratios  $R_c^{1/2}/R_a^{1/2}$ and $R_c^{1/2}/R_b^{1/2}$ reduce to
\begin{eqnarray}\label{13}
\sqrt{\frac{b_{11}(1-\tilde a_1^2-\tilde a_2^2)}{b_{33}(1-\tilde a_2^2)}}, \ \ \ \ \
  \sqrt{\frac{b_{22}(1-\tilde a_1^2-\tilde a_2^2)}{b_{33}(1-\tilde a_1^2)}}.
   \end{eqnarray}
However, these ratios determine  $F_a(\pi/2)/F(0)$, $F_b(\pi/2)/F(0)$, and so they are  equal to $29/7$ and $59/7$ (or $59/7$ and $29/7$). Thus, we have found the values of the two   combinations of the parameters (\ref{13}).

When the magnetic field rotates in the $a-b$ plane from the $a$ axis ($\phi=0$) to the direction $[110]$ ($\phi=\pi/4$), each of the frequencies $F_a(\phi)$ and $F_b(\phi)$ splits into the two branches $F_{a1}(\phi)$, $F_{a2}(\phi)$ and $F_{b1}(\phi)$, $F_{b2}(\varphi)$ if $\tilde a_2\neq 0$, i.e., if the principal axes of the ellipsoids deviate from the $a$ and $b$ axes. The $\phi$-dependences of these four branches are determined by the factors,
\begin{eqnarray} \label{14}
R_{a1,a2}(\phi)\!\!\!&=&\!\!\!b_{11}b_{33}(1-\tilde a_1^2)\sin^2\!\phi + b_{22}b_{33}(1-\tilde a_2^2)\cos^2\!\phi  \nonumber \\
&\pm& 2\sqrt{b_{11}b_{22}}b_{33}\tilde a_1\tilde a_2\sin\phi\cos\phi, \\
R_{b1,b2}(\phi)\!\!\!&=&\!\!\!b_{11}b_{33}(1-\tilde a_1^2)\cos^2\!\phi + b_{22}b_{33}(1-\tilde a_2^2)\sin^2\!\phi  \nonumber \\
&\pm& 2\sqrt{b_{11}b_{22}}b_{33}\tilde a_1\tilde a_2\sin\phi\cos\phi. \label{15}
\end{eqnarray}
At $\phi=0$, the two factors $R_{a1}(\phi)$ and $R_{a2}(\phi)$ reduce to $R_a$, whereas $R_{b1}(0)$ and $R_{b2}(0)$ coincide with $R_b$. When $\phi=\pi/4$, the four frequencies partly merge again since   $R_{a1}(\pi/4)=R_{b1}(\pi/4)$ and $R_{a2}(\pi/4)=R_{b2}(\pi/4)$. It is evident from Eqs.~(\ref{14}) and (\ref{15}) that at any $\phi$,
 \[
 R_{a1}(\phi)+R_{a2}(\phi) +R_{b1}(\phi)+R_{b2}(\phi) =2R_a +2R_b.
 \]
This equality leads to the relation between the appropriate four branches of the frequency,
\begin{eqnarray} \label{16}
\frac{1}{[F_{a1}(\phi)]^2}&+&\frac{1}{[F_{a2}(\phi)]^2}+ \frac{1}{[F_{b1}(\phi)]^2} \nonumber \\
 &+&\frac{1}{[F_{b2}(\phi)]^2}=\frac{2}{[F_{a}]^2}+\frac{2}{[F_{b}]^2},
\end{eqnarray}
where $F_a$ and $F_b$ are equal to $29$ and $59$ T.
According to Fig.~3a in Ref.~\cite{arnold16}, $F_{a1}(\pi/4)= F_{b1}(\pi/4)\approx 33$ T. This condition gives the fourth relation on the five parameters  $\tilde a_1$, $\tilde a_2$, $b_{11}$, $b_{22}$, and $b_{33}$,
\begin{eqnarray} \label{17}
\frac{[R_{c}]^{1/2}}{[R_{a1}(\pi/4)]^{1/2}}=\frac{F_{a1}(\pi/4)}{F(0)}\approx \frac{33}{7}.
\end{eqnarray}

\begin{table}
\caption{The two possible sets of the parameters specifying the W1 electrons in TaAs.}
\begin{tabular}{ccccccc}
\hline
\hline
Set~~&$\sqrt{b_{11}}$&$\sqrt{b_{22}}$&$\sqrt{b_{33}}$&~~~$\tilde a_1$~~~ &~~~$\tilde a_2$~~~&$~~~\tilde a_3$~~~\\
  &~$10^5$ m/s~&~$10^5$ m/s~&~$10^4$ m/s~& & & \\
\colrule
$1$&$3.37$&$6.74$&$6.7$&$0.5$&$0.47$&$0$\\
$2$&$6.88$&$3.30$&$6.7$&$0.5$&$0.47$&$0$ \\
\hline \hline
\end{tabular}
\end{table}

\begin{figure}[tbp] 
 \centering  \vspace{+9 pt}
\includegraphics[scale=.5]{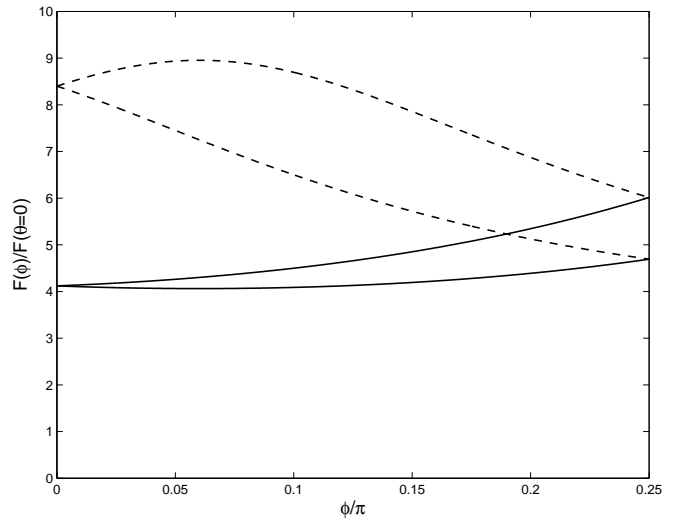}
\caption{\label{fig4} The frequencies $F_{a1,a2}$ (the solid lines) and $F_{b1,b2}$ (the dashed lines) versus the angle $\phi$ between the $a$ axis and the magnetic field lying in the $a-b$ plane. The parameters are the same as in Fig~\ref{fig2} (i.e., they coincide with the first set in Table I). }
\end{figure}

The parameter $\tilde a_1$ can be obtained from the band structure calculation along the $a$ axis. In particular, figure 1 in Ref.~\cite{arnold16} permits one to obtain the following crude estimate: $\tilde a_1\approx 0.5$. Taking into account the above four relations between $b_{11}$, $b_{22}$, $b_{33}$, $\tilde a_1$, $\tilde a_2$, we find two possible sets of the parameters characterizing the W1 points in TaAs, Table I. The dependences of the frequencies $F_{a1}$, $F_{a2}$, $F_{b1}$, $F_{b2}$ on $\phi$ for the first set of the parameters are presented in Fig.~\ref{fig4}.

At nonzero values of $\tilde a_1$ and $\tilde a_2$, the principal axes of the ellipsoid in the $a-b$ plane deviate from the $a$ and $b$ axes (Fig.~\ref{fig3}). A simple analysis leads to the following formula for the deviation angle $\psi$:
\begin{eqnarray} \label{18}
\tan(2|\psi|)=\frac{[F_{a2}(\pi/4)]^{-2}-[F_{a1}(\pi/4)]^{-2} }{[F_{a}]^{-2}-[F_{b}]^{-2}},
\end{eqnarray}
where the value of $F_{a2}(\pi/4)=F_{b2}(\pi/4)\approx 42.3$ T can be found from Eq.~(\ref{16}). Using formula (\ref{18}), we obtain $|\psi| \approx 11^{\circ}$ or $|\psi| \approx 79^{\circ}$. These two values of $|\psi|$ correspond to the two sets of the parameters in Table I. For the first set, the orientation of the ellipsoids in the $a-b$ plane is schematically shown in Fig.~\ref{fig3}. In this case, the maximal axis of their cross sections by the $a-b$ plane is inclined at the angle of $11^{\circ}$ to the $a$ and $b$ axes. For the second set, this angle is equal to $79^{\circ}$.

Knowing the angle $\psi$, the frequencies $F_1$ and $F_2$ in formula (\ref{8}) can be calculated, and we arrive at
 \begin{eqnarray} \label{19}
F_1F_2=F_aF_b\left [1-\frac{(F_{a}^{2}-F_{b}^{2})^2 }{4F_{a}^{2}F_{b}^{2}}\tan^2(2|\psi|)\right ]^{-1/2}.
 \end{eqnarray}
Eventually, expressions (\ref{7}), (\ref{8}), (\ref{19}) with $F_{3}=F(\theta=0)=7$ T  give the density $n_{W1}\approx 2.53\cdot 10^{18}$ cm$^{-3}$  produced by the eight equivalent pockets of the W1 electrons in TaAs.

\begin{thebibliography}{}


\bibitem{armit} N.P. Armitage, E.J. Mele, A. Vishwanath, Rev. Mod. Phys. {\bf 90}, 015001 (2018).

\bibitem{bernevig} A. Bernevig, H. Weng, Z. Fang, X. Dai, J. Phys. Soc. Jpn. {\bf 87}, 041001 (2018)

\bibitem{gao} H. Gao, J.W.F. Venderbos, Y. Kim, A.M. Rappe,
Annual Review of Materials Research {\bf 49}, 153 (2019).

\bibitem{wang-r} S. Wang, B.-C. Lin, A.-Q. Wang, D.-P. Yu, Z.-M. Liao, Advances in Physics: X {\bf 2}, 518 (2017).

\bibitem{m-sh19}
G.P. Mikitik, Yu.V. Sharlai, J. Low Temp. Phys.
{\bf 197}, 272 (2019)

\bibitem{zhang-nc19} C.-L. Zhang, C.M. Wang, Z. Yuan, X. Xu, G. Wang,  C.-C. Lee, L. Pi, C. Xi, H. Lin, N. Harrison, H.-Z. Lu, J. Zhang, S. Jia, Nat.\ Commun.\ {\bf 10}, 1028 (2019).

\bibitem{prl} G.P. Mikitik, Yu.V. Sharlai, Phys.\ Rev.\ Lett.\
{\bf 82}, 2147 (1999).



\bibitem{huang15} X. Huang, L. Zhao, Y. Long, P. Wang, D. Chen, Z. Yang, H. Liang, M. Xue, H. Weng, Z. Fang, X. Dai, G. Chen, Phys.\ Rev.\ X {\bf 5}, 031023 (2015). 

\bibitem{luo15} Y. Luo, N.J. Ghimire, M. Wartenbe, H. Choi, M. Neupane, R.D. McDonald, E.D. Bauer, J. Zhu, J.D. Thompson, F. Ronning,  Phys.\ Rev.\ B {\bf 92}, 205134 (2015). 

\bibitem{shek} 
C. Shekhar, A.K. Nayak, Y. Sun, M. Schmidt, M. Nicklas, I. Leermakers, U. Zeitler, Yu. Skourski, J. Wosnitza, Z. Liu, Y. Chen, W. Schneller, H. Borrmann, Yu. Grin, C. Felser, B. Yan,
Nature Physics {\bf 11}, 645 (2015).

\bibitem{arnold16} F. Arnold, M. Naumann, S.-C. Wu, Y. Sun, M. Schmidt, H. Borrmann, C. Felser, B. Yan, E. Hassinger, Phys.\ Rev.\ Lett. {\bf 117}, 146401 (2016). 

\bibitem{hu16} J. Hu, J.Y. Liu, D. Graf, S.M.A. Radmanesh, D.J. Adams, A. Chuang, Y. Wang, I. Chiorescu, J. Wei, L. Spinu, Z.Q. Mao,  Sci. Rep. {\bf 6}, 18674 (2016). 

\bibitem{du16} J. Du, H. Wang, Q. Chen, Q. Mao, R. Khan, B. Xu, Y. Zhou, Y. Zhang, J. Yang, B. Chen, C. Feng, M. Fang, Sci. China-Phys. Mech. Astron. {\bf 59}, 657406 (2016).

\bibitem{sergelius} P. Sergelius   
et al., Sci. Rep. {\bf 6}, 33859 (2016).

\bibitem{wang16} Z. Wang, Y. Zheng, Z. Shen, Y. Lu, H. Fang, F. Sheng, Y. Zhou, X. Yang, Y. Li, C. Feng, Z.-A. Xu, Phys.\ Rev.\ B {\bf 93}, 121112 (2016). 



\bibitem{he-r(h)} 
L.P. He, X.C. Hong, J.K. Dong, J. Pan, Z. Zhang, J. Zhang, S.Y. Li, Phys. Rev. Lett. {\bf 113}, 246402 (2014).

\bibitem{pari} A. Pariari, P. Dutta, P. Mandal, Phys.\ Rev.\ B {\bf 91}, 155139 (2015). 

\bibitem{liang}
T. Liang, Q. Gibson, M.N. Ali, M. Liu, R.J. Cava, N.P. Ong,
Nature Materials {\bf 14}, 280 (2015). 

\bibitem{zhao15} Y. Zhao, H. Liu, C. Zhang, H. Wang, J. Wang, Z. Lin, Y. Xing, H. Lu, J. Liu, Y. Wang, S.M. Brombosz, Z. Xiao, S. Jia, X.C. Xie, J. Wang, Phys.\ Rev.\ X {\bf 5}, 031037 (2015). 

\bibitem{nara15} A. Narayanan, M.D. Watson, S.F. Blake, N. Bruyant, L. Drigo, Y.L. Chen, D. Prabhakaran, B. Yan, C. Felser, T. Kong, P.C. Canfield, A.I. Coldea, Phys.\ Rev.\ lett. {\bf 114}, 117201 (2015).

\bibitem{xiang15} Z.J. Xiang, D. Zhao, Z. Jin, C. Shang, L.K. Ma, G.J. Ye, B. Lei, T. Wu, Z.C. Xia, X.H. Chen, Phys.\ Rev.\ lett. {\bf 115}, 226401 (2015). 
    
\bibitem{desr} W. Desrat, C. Consejo, F. Teppe, S. Contreras, M. Marcinkiewicz, W. Knap, A. Nateprov, E. Arushanov, J. of Phys.:Conf. Ser. {\bf 647}, 012064 (2015). 

\bibitem{cao15} J. Cao, S. Liang, C. Zhang, Y. Liu, J. Huang, Z. Jin, Z.-G. Chen, Z. Wang, Q. Wang, J. Zhao, S. Li, X. Dai, J. Zou, Z. Xia, L. Li, F. Xiu, Nat. Commun. {\bf 6}, 7779 (2015).   

\bibitem{he16} L.-P. He, S.-Y. Li, Chin. Phys. B {\bf 25} 117105 (2016). 

\bibitem{crassee} I. Crassee, R. Sankar, W.-L. Lee, A. Akrap, M. Orlita, Phys.\ Rev.\ Materials {\bf 2}, 120302 (2018).
    



\bibitem{m-sv} G.P. Mikitik, I.V. Svechkarev, Fiz. Nizk. Temp.
{\bf 15}, 295 (1989) [Sov. J. Low Temp. Phys. {\bf 15}, 165 (1989)].
See also: http://www.ilt.kharkov.ua/bvi/structure/depart\_e/d26/\\
publ\_mik\_shar/11\_en.pdf

\bibitem{m-sh} G.P. Mikitik, Yu.V. Sharlai, Fiz. Nizk. Temp.
{\bf 22}, 762 (1996) [Low Temp. Phys. {\bf 22}, 585 (1996)].
See also: http://www.ilt.kharkov.ua/bvi/structure/depart\_e/d26/
\\publ\_mik\_shar/18\_en.pdf

\bibitem{sol} A. Soluyanov, D. Gresch, Z. Wang, Q. Wu, M. Troyer, X. Dai, and B.A. Bernevig, Nature {\bf 527}, 495 (2015). 


\end{thebibliography}
\end{document}